\documentclass[10pt]{article}          % LaTeX 2e 
\usepackage{times}                               % LaTeX 2e 
\usepackage{latexsym}                               % LaTeX 2e 
\usepackage{amssymb}
\usepackage[pdftex]{graphicx}
\usepackage{filemacro2}
\usepackage{color}
\begin{document}

%\mainmatter  % start of an individual contribution

\title{Improving Memory Hierarchy Utilisation for Stencil Computations on Multicore Machines\footnote{Supported by CNPq and partially supported by the Brazilian Oil Company (Petrobras)}}

\author{Alexandre Sena \\
         {\small Institutos Superiores de Ensino La Salle, Niter\'oi, RJ, Brazil }\\ \\
        Aline Nascimento, Cristina Boeres, Vinod E.\ F.\ Rebello  \\
        {\small \{aline,boeres,vinod\/\}@\/ic.uff.br }\\
        {\small Instituto de Computa\c c\~ao, Universidade Federal Fluminense (UFF),  Brazil }\\\\
        Andr\'e Bulc\~ao \\
        {\small  bulcao@petrobras.com.br }\\
        {\small Centro de Pesquisas e Desenvolvimento Leopoldo Am\'erico Miguez de Mello, Petrobras, RJ, Brasil }\\
 }

\maketitle

\begin{abstract}

Although modern  supercomputers are composed of multicore machines, one can find scientists that still execute their legacy applications which were developed to  monocore cluster where memory hierarchy is dedicated to a sole core. The main objective of this paper is to propose and evaluate an algorithm that identify an efficient blocksize to be applied on MPI stencil computations on multicore machines. Under the light of an extensive experimental analysis, this work shows the benefits of identifying blocksizes that will dividing data on the various cores  and suggest a methodology that explore the memory hierarchy available in modern machines.

\end{abstract}

%-------------------------------------------------------------------------
\section{Introduction}

Despite the increase in processing power and storage capacity of computers in recent decades~\cite{JimGray}, there is a growing demand by scientists, engineers, economists among others researchers, which are looking for high performance computing. This demand is due to a variety of complex problems being studied and also the increasing amount of data to be analysed. 

 In modern system platforms, a growing number of multiprocessors~\cite{clapp2010} has become available and researchers have been spending significant efforts to extract high performance from such systems~\cite{icl:658}. In recent studies, the memory access pattern is a crucial aspect when designing a 3D domain problem, as in the case of the stencil computations on regular grid~\cite{Kamil:2005}, which appears in many problems of different areas.

%For example,  versions of the kernel of the RTM code have been implemented not only on conventional CPU multicore processors, but also on FPGAs (Field Programmable Gate Arrays)~\cite{seg2008} and GPGPU (General Processing Graphics Processing Units)~\cite{Micikevicius:2009}. General CPU systems are the most flexible, allowing a variety of optimisations and algorithmic manipulations. Furthermore, the majority of the RTM programs available are implemented in FORTRAN or C, using MPI or OpenMP, and to be readily portable to these omnipresent high performance platforms. 

%However, while the new implementations for GPGPU and FPGA tries to extract the best performance of these architectures, 

In general, the CPU legacy codes are directed executed on multicore machines (normally, only a recompilation is necessary), without being tuned to the peculiarities of these multicore architectures. In this case, important features, as the cache memory hierarchy that are shared between different cores, are ignored. Thereby, efforts are necessary to improve the utilisation of cache memory, in order to increase performance. 

An effective approach to diminish the memory access cost by increasing data locality is the technique called blocksize calculation or Tiling
~\cite{Abella02,seg2008ortigosa,Rivera2000,IPDPSDongarra2012}, where the loops are separated into smaller ranges so that data will remain in the cache while required. The difficulty lies on identifying the block value that will lead to the smallest execution time.

The main objective of this work is to address the fact that when executing stencil application on multicore machines, performance is gained if  memory hierarchy is well explored. An extensive analysis on the performance of a typical  strategy that identifies efficient blocksize or tiling to be used on stencil computations is carried out. Based on this study, a blocksize specification methodology is suggested and a detailed evaluation shows its benefits when executing stencil application on multicore machines. Through an extensive experimental evaluation, this work suggests a simple way for scientists to identify an efficient blocksize, shown to be very close to the optimal one.

% Interesting Comments
%  One solution to this problem is to hand-tune the relevant computational kernel for a given platform; this would allow the programmer to specify the domain-specific transformations that the compiler alone was unable to exploit

% we show that that the best code and runtime parameters for one multicore machine often do not correspond to good performance on other multicore machines. Therefore, the methodology to find an efficient blocksize should be executed whenever a new system is acquired. --> auto-tunning? not really

% Should we not address auto-tunners? auto-tuning has already had several previous success stories, including: FFTW [20], SPIRAL [41], OSKI [54], and ATLAS [55]

%-------------------------------------------------------------------------
\section{3D Stencil Computations}

\begin{figure}[htb]
\centering
\includegraphics[scale=0.33]{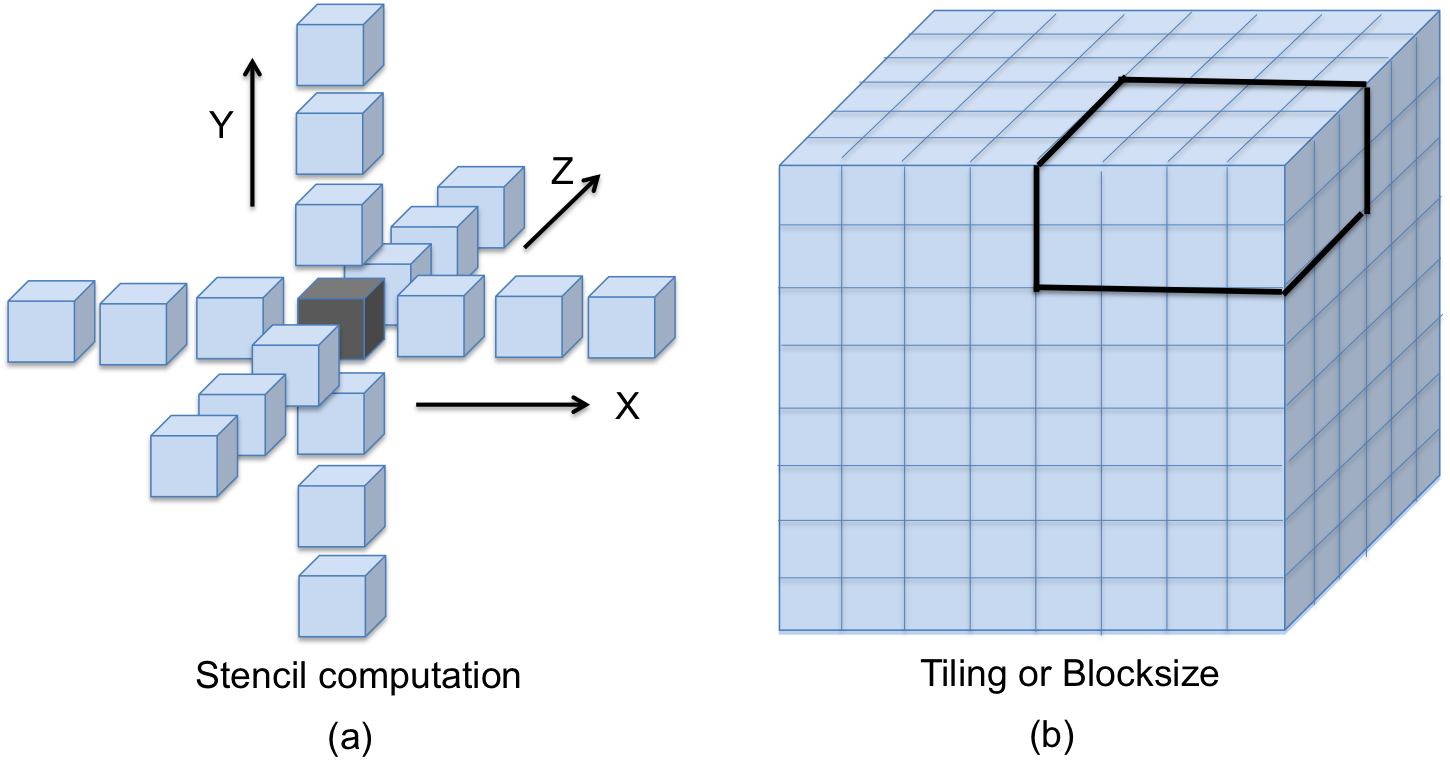}
\caption{\small (a) Stencil computation of the dark point. (b) A blocksize for a 3D grid domain.}
\label{stencil}
\end{figure}

Partial differential equation solvers  are often implemented using iterative finite-dif\-fe\-ren\-ce techniques that sweep over a spatial grid, performing nearest neighbour computations called stencils.  These computations are employed by a  variety of scientific applications in   heat diffusion, electromagnetism, fluid dynamics and seismic imaging.  In a stencil operation, each point in a regular grid is updated with weighted contributions from a subset of its neighbours in both time and space~\cite{Kamil:2005}.  As can be seen in Figure~\ref{stencil}(a), a new value for the dark cell  is obtained by the weighted sum of the values in all three neighbour cells in all three dimensions in both directions.

More specifically, in this work the behaviour of the stencil computation on the RTM application~\cite{seg2008ortigosa,Araya-Polo:2009} will be studied. The RTM is used to produce accurate imaging of new oil fields discovered in the Gulf of Mexico and in Brazil's southeast coast  located in deep water, despite its high computational cost. The most time consuming part of the RTM algorithm is the computation of the 3D model, as indicated in line~\ref{model} of the RTM Algorithm~\ref{alg:rtm}, which is based on the difference finite method for PDE of the $10^{th}$ order in space and $2^{nd}$ order in time. This 3D method has a memory access pattern that refers to five points (cells) in each direction for each dimension to calculate the value of the actual point and actually, corresponds to 99\% of the total  execution time of the RTM application, as evaluated in~\cite{eScience2011}.

In general,  stencil calculations perform global sweeps through data structures that are typically much larger than the cache capacity, and also, data reuse is limited to number of neighbour points. Thus, optimisations are necessary to take full advantage of the memory hierarchy and to achieve high performance. The new multicore processors have some in-core optimisations to enhance performance, as for example, the Intel Streaming SIMD Extensions (SSE) allows data parallelism to execute the same operation on distinct elements in a data set~\cite{stencilIntel}. However, the data parallelism will not be a significant help if the stencil computation underutilize cache lines. Data locality is the main target to achieve high performance in stencil computations.

On 3D stencil applications only one dimension has sequential data in memory which permits a fast execution, specially when vectorization is applied. On the other hand, the memory accesses  considering the remaining directions are expensive. Therefore, some efforts are necessary to improve the data locality on such computations. One approach is to apply transformations to improve the cache memory utilisation. Loop splitting is an optimisation that breaks a loop into two or more loops, each one with fewer computations to reduce register utilisation. By splitting the stencil computation on independent $x$, $y$ and $z$ directions, the number of SIMD registers required to parallelise each loop is considerable smaller than the original one single whole loop~\cite{stencilIntel}. Another effective approach is the technique called Tiling or Blocksize that will be explained in the next section.

\vspace{0.2cm}

\begin{center}
\parbox{80mm}{
\begin{algorithm}{Algorithm}{\sf RTM ()}
\label{alg:rtm}
\algline{initialisation; }
   \algline{for each time step \{}
    \algindent
    	\algline{{forall}  points: calculate energy; \label{energy}}
     	\algline{{forall}  points: calculate sismic; \label{sismic}}
   	 \algline{{forall}  points:  calculate 3D model;  \label{model}}
   	 \algreturn{\}}
\end{algorithm}
}
\end{center}

\section{Blocksize Technique for 3D Stencil Algorithms}

The memory access pattern is a crucial aspect when designing a 3D domain problem, as in the case of the RTM kernel program~\cite{Kamil:2005}.
An effective approach to diminish the memory access cost by increasing data locality is the technique called blocksize calculation or Tiling
~\cite{Abella02,seg2008ortigosa,Rivera2000}, where the loops are separated into smaller ranges so that data will remain in the cache while required. The difficulty lies in identifying the block size value that will lead to the smallest execution time.  Figure~\ref{stencil}(b) shows a block of elements to be traversed considering that its size has been calculated.

As a first step, this work shows a series of experiments that will portray the effect of the block size value on the application performance, focusing on  stencil applications. For a better understanding of the performance, the Intel Vtune application profiler~\cite{VTune} is used. 

%%%%%%%%%%%%%%% aqui??  %%%%%%%%%%

\subsection{Blocksize Profiler Analysis}

Three versions of the sequential 3D model code for the problem size of $200 \times 200 \times 800$ was sampled in a node with two Intel Xeon E5410 2.33GHz Quad core processors with 12MB L2 Cache and 16GB of RAM memory per node.  The only difference between the three versions analysed is the blocksize used. The first version, denoted as  {\sf No-Blocksize}, no blocksize technique was applied, while in a second version, called  {\sf Wrong-Blocksize}, inefficient blocksize was given ($159 \times 209 \times 175$) and finally, in the third version, denoted as  {\sf Efficient-Blocksize}, the application was executed with an efficient blocksize ($15 \times 15 \times 143$).

For this analysis, we consider a sample period of 50 iterations instead of the original 448 iterations in the remaining experiments evaluated in this paper. 
Table~\ref{vtune1} presents the number of events that were collected for the three versions for a dedicated execution on only one unique core.

%TABELA VTUNE  1 CORE 
%\vspace{-0.3cm}
\begin{table}[htb]
\begin{small}
\begin{center}
\caption{Number of events collected by Vtune tool when executing one process of the application on one core}
\begin{tabular}{|l|r|r|r|} \hline
%{code modules} & {}    & \multicolumn{3}{c|}{{\it MWork MPI}} & \multicolumn{1}{c|}{{\it MWork}}\\ \cline{3-6}
{Events in millions ($\times 10^6$)}                        & { {\sf No-Blocksize} }     & { {\sf Wrong-Blocksize} }  & { {\sf Efficient-Blocksize} }      \\ \hline \hline
{Clockticks}                                 &    {116,131}   & {61,984}    & {53,599}      \\ \hline
{Retired instructions}                 &    {56,845}   & {63,445}    & {65,561}     \\ \hline
{L1 misses}                                 &    {1,275}    & {1,013}     & {1,026}      \\ \hline
{Number of lines in L2}             &    {900}    & {628}      & {318}       \\ \hline
{L2 misses}                                 &    {380}   & {46}     & {14}         \\ \hline
{Number of cycles with stall}    &    {83,032}   & {28,873}    & {20,609}     \\ \hline
{Number of bus cycles}              &    {16,522}   & {8,846}     & {7,762}      \\ \hline
{Number of bus transactions}       &    {1,899}    & {1,359}      & {730}         \\ \hline

\end{tabular}
\label{vtune1}
\end{center}
\end{small}
\end{table}
%\vspace{-0.3cm}

  {\sf Efficient-Blocksize} achieved the best performance, as justified by the smallest number of clockticks, since   {\sf Wrong-Blocksize} and   {\sf No-Blocksize} obtained values that were 15.6\% and 116.6\% larger, respectively. Interestingly,   {\sf Efficient-Blocksize} produced more retired instructions than any other version, showing that the effectiveness of the code permitted the processor to retire more than one instruction per cycle, producing the smallest number of clockticks.

Concerning memory issues,  {\sf Efficient-Blocksize} also performed better than the other versions, specially when evaluating L2 cache related issues.   {\sf Wrong-Blocksize} and   {\sf No-Blocksize} produced 228.5\% and 2,614.2\%, respectively, more L2 cache misses than  {\sf Efficient Blocksize}. One can note that the  {\sf Wrong-Blocksize} and  {\sf No-Blocksize} allocated 97.4\% and 183\% more L2 cache lines than  {\sf Efficient-Blocksize}. Based on this event, the L2 cache miss rate derived for each version
($\frac{\mbox{\it \small Number of lines in L2}}{\mbox{\it \small Retired Instructions}}$) is 0.005, 0.010 and 0.016 for  {\sf Efficient-Blocksize}, {\sf Wrong-Blocksize} and  {\sf No-Blocksize}, respectively. Actually, the cache miss rate for {\sf Efficient-Blocksize} is an order of magnitude smaller than the other versions, what is reflected directly on the performance, showing the benefits of avoiding misses on higher memory levels.

Another event that portray memory access is the utilisation of the system memory bus (FSB), which connects the processor with RAM memory and can indicate performance bottlenecks. As can be seen in Table~\ref{vtune1}, the number of bus cycles is smaller for  {\sf Efficient-Blocksize}, meaning that a smaller number of accesses to external memory
was necessary due to better data locality. The bus utilisation metric ($\frac{\mbox{\it \small Number of bus transactions}\times 2}{\mbox{\it \small Number of bus cycles}}$) for versions {\sf Efficient-Blocksize}, {\sf Wrong-Blocksize} and  {\sf No-Blocksize} are 0.19, 0.31 and 0.23, respectively. 

Finally, the smallest number of clock cycles that stalls due to branch misprediction was  for the  {\sf Efficient-Blocksize} version, while {\sf Wrong-Blocksize} and  {\sf No-Blocksize} were 40\% and 302\% larger than it. The resource stall ratio ($\frac{\mbox{\it \small Number of cycles with stall}}{clockticks}$)  confirms the better performance of version {\sf Efficient-Blocksize} being only 38\% and smaller than the other versions.

This same experiment was repeated for the three versions, however, instead on a dedicated node, the profiling of each version was
executed together with another seven processes of the same version on the remaining cores of the node. This analysis is more representative in the sense that it reflects the actual memory utilisation of the execution of the RTM application (stencil computation) in multicore machines, when one process per core is executed. The results obtained  are presented in Table~\ref{vtune2}.

%TABELA VTUNE  8 CORES (COMPARTILHANDO RECURSOS) 
%\vspace{-0.3cm}
\begin{table}[htb]
\begin{small}
\begin{center}
\caption{Vtune events executing on all 8 cores of a node}
\begin{tabular}{|l|r|r|r|} \hline
%{code modules} & {}    & \multicolumn{3}{c|}{{\it MWork MPI}} & \multicolumn{1}{c|}{{\it MWork}}\\ \cline{3-6}
{Events in millions ($\times 10^6$)}                        & { {\sf No-Blocksize} }     & { {\sf Wrong-Blocksize} }  & { {\sf Efficient-Blocksize} }      \\ \hline \hline
{Clockticks}                                     &    {1,942,215}   & {1,009,088}     & {615,907}      \\ \hline
{Retired instructions}                    &    {405,280}       & {505,356}        & {468,642}     \\ \hline
{L1 misses}                                     &    {10,229}         & {8,309}             & {7,488}      \\ \hline
{Number of lines in L2}                 &    {7,565}           & {4,227}             & {2,471}       \\ \hline
{L2 misses}                                     &    {5,259}           & {1,497}             & {621}         \\ \hline
{Number of cicles with stall}        &    {1,552,245}   &  {706,546}        & {365,191}     \\ \hline
{Number of bus cicles}                 &    {253,593}       & {124,845}        & {88,002}      \\ \hline
{Number of bus transactions}       &    {86,573}       &  {46,363}          & {31,754}         \\ \hline

\end{tabular}
\label{vtune2}
\end{center}
\end{small}
\end{table}
%\vspace{-0.3cm}

The results confirmed the advantages of the use of an efficient blocksize. For instance, the number of clock ticks produced by  {\sf Wrong-Blocksize} and  {\sf No-Blocksize} are, respectively, 63.8\% and 215.3\% larger than those of {\sf Efficient-Blocksize}. Also, the number of L2 cache misses were  more than twice smaller than  {\sf Wrong-Blocksize} and more than eight times smaller than {\sf No-Blocksize}. When considering the clock cycles that stalls, the versions {\sf Wrong-Blocksize} and {\sf No-Blocksize} presented, respectively, 93,4\% and 325\% more than {\sf Efficient-Blocksize}. Besides, the resource stall ratio for  {\sf Efficient-Blocksize} was the smallest amongst  the three versions and the only one below 60\%, which is considered the threshold for a efficient execution.

Finally, to evaluate the execution time of the application for the three cases, they were executed on one dedicated core of a node and also, on the eight cores of the node, one process per core, sharing the memory hierarchy. In both executions,  each process computeed 448 iterations for the $200 \times 200 \times 800$ problem size instance. The results can be seen in Table~\ref{ExecTime3}.

%TABELA VTUNE  8 CORES (COMPARTILHANDO RECURSOS) 
%\vspace{-0.3cm}
\begin{table}[htb]
\begin{small}
\begin{center}
\caption{Execution time of the three versions (in seconds)}
\begin{tabular}{|c|c|c|c|} \hline
%{code modules} & {}    & \multicolumn{3}{c|}{{\it MWork MPI}} & \multicolumn{1}{c|}{{\it MWork}}\\ \cline{3-6}
{Number of cores}                        & { {\sf No-Blocksize} }     & { {\sf Wrong-Blocksize} }  & { {\sf Efficient-Blocksize} }      \\ \hline \hline
{1}                    &    {390.0}   & {279.1}     &  {234.3}      \\ \hline
{8}                    &    {1033.4}   & {825.2}     &  {488.1}     \\ \hline
\end{tabular}
\label{ExecTime3}
\end{center}
\end{small}
\end{table}
%\vspace{-0.3cm}

As was demonstrated by the VTune analysis, the performance of the version with efficient blocksize is clearly the best. When considering the execution in a dedicated node to the sole process the execution time of  {\sf Wrong-Blocksize} and {\sf No-Blocksize} were 19.1\% and 66.5\% slower than the  efficient blocksize one. More important, when considering the sharing of the resources of a node with other processes, the performance of {\sf Efficient-Blocksize} were even more impressive, being the execution time of the  {\sf Wrong-Blocksize} and {\sf No-Blocksize} 69\% and 111.7\% slower than it. 

Note that, when executing with other processes sharing the available resources, all three versions should produce execution times close to the ones obtained by their respective execution on one core, since  the problem size executed per process is the same. However, as it was analysed with  the VTune tool, the sharing of the memory hierarchy has a deep impact in the execution, diminishing the performance of the application. Comparing the execution time on one core with the one on eight cores,   an increase of 108\%, for {\sf Efficient-Blocksize} was detected. On the other hand, the two other versions were even worse, producing execution times on eight cores three times higher than on one core.

These two experiments definitely presented the advantages of an appropriated blocksize and, its benefits  are even more striking if we consider larger problem sizes and an increase in the number of iterations. Thus, the next section presents an alternative approach to find efficient blocksizes in multicore machines.

%%%%%%%%%%%%%%%%%%%%%%%%%%%%%%

\subsection{A Framework for Profiling Technique}
\label{s:framework}

In order to identify the best value for the block size, a typical technique is dynamic profiling of the code under consideration, which is the division of the work among all the processing units available, during which each processing core calculates the execution time of a given number of  iterations of the kernel with distinct blocksizes. However, even for a small 3D problem size, testing all possible coordinates combinations is practically unfeasible. The approach as described in~\cite{eScience2011}, specifies  $nC$  combinations of  points $x$, $y$, and $z$ (a block of elements to be transversed is then defined), which are first tested by executing a number $nI$ of iterations of the program kernel with different blocksizes (combinations) and chooses the block value associated with the smallest execution time.

The framework for profiling adopted in this work  implements two distinct phases, the {\em selection} and {\em verification} phases, as described as follows:
\begin{description}

\item[\em Selection Phase: ] a fixed number of $nC$ block values combinations  are executed by the $nP$  processes in accordance with a given distribution on the processing elements of the target system and evaluated in parallel. Each one of the $nP$ processes executes only $nI$ iterations of the RTM Kernel. The chosen blocksize is the one  associated with a given criteria that optimises the execution time of the $nI$ iterations of the kernel. This best execution time will be denoted as $MinTime$.  

\item[\em Verification Phase: ]  each one of the $nP$ processes executes  $nI$ iterations of the kernel altogether with the given best block value found during the selection phase. The  execution time measured at this phase will be denoted $ActualTime$. 

\end{description}

\subsection{Approaches to identify the best blocksize}
 \label{novoAlg}

 The two phase profiling framework was implemented on the so called Original Block (OB) methodology, where during the {\em selection phase}, each one of the $nP$ processes executes $\frac{nC}{nP}$ blocksize combinations in any order. The blocksize with the minimal execution time is the chosen one.

Nevertheless, when the processes execute distinct blocksize combinations asynchronously during the selection phase, different amounts of cache memory might be  allocated to each process, which can lead to a misleading evaluation.  The main idea of the new approache is to force all  cores of each node to evaluate the same block value combinations in similar order, so that  the same amount of cache memory is allocated to each one of  the processes of a given node  (i.e. the cache memory is equally divided among the cores of a node). 

This methodology is defined as the following. Let $N$ be the total number of multicore nodes and let $NC_i$ be the set of cores of node $i$. During the {\em selection phase}, the $nC$ combinations are equally divided between the nodes and then, each one of the $\mid NC_i \mid$ cores of a each node evaluates $\frac{nC}{N}$ combinations in the same order, in three stages:

\begin{description}
\item[Stage 1:] selection of the best blocksize of each core;
\item[Stage 2:] selection of the best blocksize of each node;
\item[Stage 3:] selection of the best blocksize among all nodes. 
\end{description}

\begin{figure}[htb]
\centering
\includegraphics[scale=0.5]{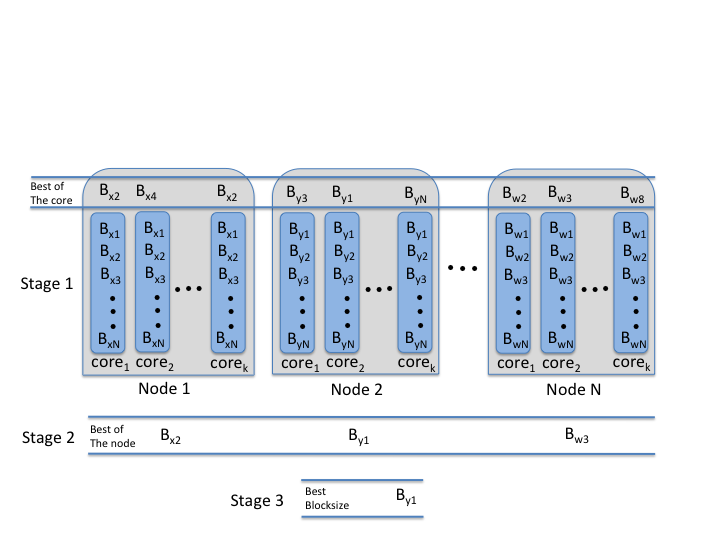}
\caption{\small The three stages of the new approach for multicore machines}
\label{FigAlgNew}
\end{figure}

A visualisation of the execution of the  three stages can be seen in Figure~\ref{FigAlgNew}, where, in each core, a sequence of block values are evaluated and the best one is chosen.

A natural approach is to choose the blocksize that minimises the execution time at each one of the three stages. For the first stage, this choice is the best one because the blocksize is chosen among completely distinct block values and  the shared memory space is not considered. For the third stage, this choice is also the best one because the chosen one is evaluated considering distinct values on a non-shared memory space. This is not true for the second stage, during which although all the processes of a node (hopefully) evaluate the same blocksizes in the same order, a minimisation function would not necessarily lead to the best value. This is due to the fact that there is no guarantee that the choice occurs on distinct values. Therefore, in this new algorithm, the first stage minimises its objective function,  i.e.\ it chooses the blocksize with the smallest execution time since this is done sequentially by the process on its core. In the same manner, the third stage also minimises its objective function, since at this point, no cache influence occurs (data is allocated on the distributed memory). On the other hand, during the second stage, three different approaches were evaluate: the choice of  the block size that leads to the minimum, maximum and average executions time.

More formally, let $P_{k,i}$ be the process executing on core $c_{k,i} \in NC_i$ in node $i$. The time for $nI$ iterations of the kernel to be executed by $P_{k,i}$ using block value $b_k$ is given by $t(b_k, P_{k,i} )$. In the first stage,  each $P_{k,i}$ selects the block value $b^{min}_{k,i}$  such that $t( b^{min}_{k,i}, P_{k,i}) = \min_{\forall k} t( b_k, P_{k,i})$. 
Thus, as seen in Figure~\ref{FigAlgNew}, at the end of this stage, each processor $P_{k,i}$ will have chosen a block value $b^{min}_{k,i}$, resulting in a total of $\sum_{1}^{N} NC_i$ blocksizes. 

Then, in the second stage, the best block value amongst  the cores in $NC_i$  in node $i$ considering each evaluated function is given by: 

\begin{equation}
\label{eq_min}
B^{min}_{NC_i} =  \min_{P_{k,i} \in NC_i} \{ t(b^{min}_{k,i}, P_{k,i}) \} 
\end{equation}
\begin{equation}
\label{eq_max}
B^{max}_{NC_i}  =  \max_{P_{k,i} \in NC_i} \{ t(b^{min}_{k,i}, P_{k,i})  \}  
\end{equation}
\begin{equation}
\label{eq_avg}
B^{avg}_{NC_i} =   \min_{P_{k,i} \in NC_i} \{ | t( b^{min}_{k,i}, P_{k,i}) - \frac {\sum_{j=1}^{i}t( b^{min}_{k,i}, P_{k,i})}{i}| \}  
\end{equation}

At the end of the second stage, there will be $N$ block values choses, one for each node $i$. 
Finally, in the third stage, the best block value $b_{best}$ is such that 
\[ b_{best} = b_k \mid t(b_k, p ) = \min_{\forall N_i} t( B^{target}_{N_i}, p_j) \]
 is the one that provided the smallest execution time amongst all blocksize chosen for each node, where $target$ is either $min$, $max$ or $avg$, respectively. The mechanisms that calculate  $b_{best}$  based on $B^{min}_{N_i}$, $B^{max}_{N_i} $ and $B^{avg}_{N_i} $  are denoted as Min-Min-Min  ($MMMB$), Min-Worst-Min Block ($MWMB$) and Min-Average-Min Block ($MAMB$), respectively.

\subsection{Comparing the Original and Proposed Methodologies}

Aiming to analyse the benefits and pitfalls of the strategies presented, a first series of experiments were carried out on a multicore cluster, called Oscar,  composed of 40 BULL nodes, interconnected by Gigabit Ethernet, each node with two Intel Xeon E5430 2.66 GHz Quad core processors with 12MB L2 Cache and 16GB of RAM memory per node, running RHEL 5.3 and NFS.  

An extensive analysis on three problem sizes: $200 \times 200 \times 800$, $300 \times 300 \times 800$ and $250 \times 250 \times 1500$, representing small, medium and large instances concerning the use of RAM memory, was reported in~\cite{eScience2011} and highlighted the fact that  the execution times for the chosen blocksizes were not uniform, exhibiting a difference of 92\% for problem size $300 \times 300 \times 800$ and of 96\%  for problem size $250 \times 250 \times 1500$, when comparing the smallest and largest execution times.

The difference between $MinTime$ and  $ActualTime$ times as defined in Section~\ref{s:framework}, also called one's attention, since the same block values were used. These  results showed that  $OB$ did not produce an adequate block value that maximises the cache utilisation on the target multicore systems.  More specifically, let $P_i$ and $P_j$  be two processes executing the selection phase of the algorithm in the same node $N_k$ with two cores that share the cache memory. It is probable that $P_i$ and $P_j$ will execute with different blocksize values, such that distinct amounts of memory, for instance $m_i$ and $m_j$, will be necessary. The size of cache shared by $P_i$ and $P_j$ reflects in the computational time of $P_i$ and  $P_j$.
Let $P_i$ be the process that identified the blocksize that produced the minimum time $t_i$. In the verification phase the processes will
execute the code with the same block value, with an amount of memory $m_i$. However, in this execution, the actual execution time $t_i'$ can
be greater than $t_i$, because the amount of cache memory used by the two processes will be $2 \times m_i$ instead of $m_i + m_j$. This
behaviour motivates the creation of the strategy proposed in the previous section, which produces a blocksize associated with a minimum time in the selection phase close to the actual execution time during the verification phase.

The proposed mechanisms were also executed on $128$ cores of Oscar cluster for the three distinct problem sizes. The results provided were more homogeneous than $OB$, considering the 10 executions of each instance. Table~\ref{tabResumo} summarises these results (in seconds) where one can see  the best, average and worst times considering the 10 executions associated with each algorithm and also the respective standard deviation (ST) for each problem size.

Table~\ref{tabResumo} shows that $MWMB$ obtained  the best results practically for all the instances analysed, losing only once (third line) for   $MAMB$, but  with a difference of less than 1.8\%. Also, $MWMB$ were significantly more efficient than  $OB$ and $MMMB$. Another important advantage of the $MWMB$ and $MAMB$ algorithms are the similarity of the block values calculated in all of the 10 executions. For all of the problem sizes, the standard deviation (ST) was very small, showing the accuracy and efficiency of the $MWMB$ and $MAMB$ algorithms. 
The main reason for such good results is that this technique provides a more fair division of the cache memory among the cores of a multicore node. A consequence of this efficiency when comparing the differences  of $MinTime$ and  $ActualTime$ times obtained by the selection and verification phases, respectively, being  always less than 1\% (Table~\ref{minActualTimes}, to be analysed later).

Therefore, from now on, the algorithm $MWMB$ will be adopted as the basis of the study carried out in the remaining of this paper.

\begin{table}[htb]
\caption{Comparison of the four procedures}\label{tabResumo}
  \centering
  \begin{small}
\begin{tabular}{|c|c|c|c|c|c|}
  \hline
  % after \\: \hline or \cline{col1-col2} \cline{col3-col4} ...
  Results & problem size       & OB     & MMMB   &   MWMB &  MAMB      \\  \hline
  {} & $200 \times 200 \times 800$         & 914.3  & 700.3       &   {\bf 621.0}    & 678.0   \\ \cline{2-6}
  {best} & $300 \times 300 \times 800$     & 1279.4    & 1418.0     &   {\bf 1278.2}  &   1411.5 \\ \cline{2-6}
  {} & $250 \times 250 \times 1500$        & 1926.5  & 2107.2      &   1891.1      &  {\bf 1857.6} \\  \hline
  {} & $200 \times 200 \times 800$         & 963.2  & 906.4       &   {\bf 631.8}    &   688.5   \\ \cline{2-6}
  {average} & $300 \times 300 \times 800$  & 1648.2    & 1537.0     &   {\bf 1295.0}  & 1443.8 \\ \cline{2-6}
  {} & $250 \times 250 \times 1500$        & 3281.0  & 2439.7      &  {\bf 1926.1}   &   2043.7 \\ \hline
  {} & $200 \times 200 \times 800$         & 1145.4  & 981.0       &   {\bf 655.3}     &   698.0 \\ \cline{2-6}
  {worst} & $300 \times 300 \times 800$    & 2467.5  & 2451.4     &   {\bf 1313.2}  &  1693.7\\ \cline{2-6}
  {} & $250 \times 250 \times 1500$        & 3674.6  & 3615.0      &  {\bf 1958.3}     & 2135.0\\ \hline\hline
   {} &  $200 \times 200 \times 800$       & 68.2      & 107.7       &   10.2    &   6.4 \\ \cline{2-6}
   {ST} & $300 \times 300 \times 800$   & 540.5    & 321.6      &   11.2     &    87.9 \\ \cline{2-6}
  {} & $250 \times 250 \times 1500$       & 713.3    & 613.2      &  20.2   &   120.6 \\ \hline

\end{tabular}
\end{small}
\end{table}

%Not always choosing the Blocking that minimises the execution time can be a good strategy, specially because the best execution time produced by the algorithm can reflect only the better cache memory  of the process that obtained the minimum time and not because a better utilisation of all processes that share the cache memory. As can be clearly seen in the results, the algorithm $MWMB$ that has a maximazion stage that get the blocksize that produces the worse execution time from all blocksizes chosen by all cores of a same node was more effective than the others obtained, in general, better performance.  

\subsection{The quality of the selected blocksize }
\label{quality}

The next experiment exhaustively executes the RTM kernel for a different set of blocksize values in order to obtain the one with the smallest execution time. Comparing this blocksize value with the one produce by $MWMB$  provides an opportunity to measure the quality of the proposed strategy.

When executing the RTM kernel for a $250 \times 250 \times 1500$ problem size instance per process, a total of  $250 \times 250 \times 1500 \cong 94 $ millions block combinations  would be necessary to attain the best blocking value. In order to reduce the number of tests and thus carry out this experiment in a feasible time, the blocking values were tested at intervals of 16 points for $i$ and $j$, and 32 points for $k$. This experiment was carried out on one eight cores Oscar machine. 

\begin{figure}[h]
\centering
\includegraphics[scale=0.37]{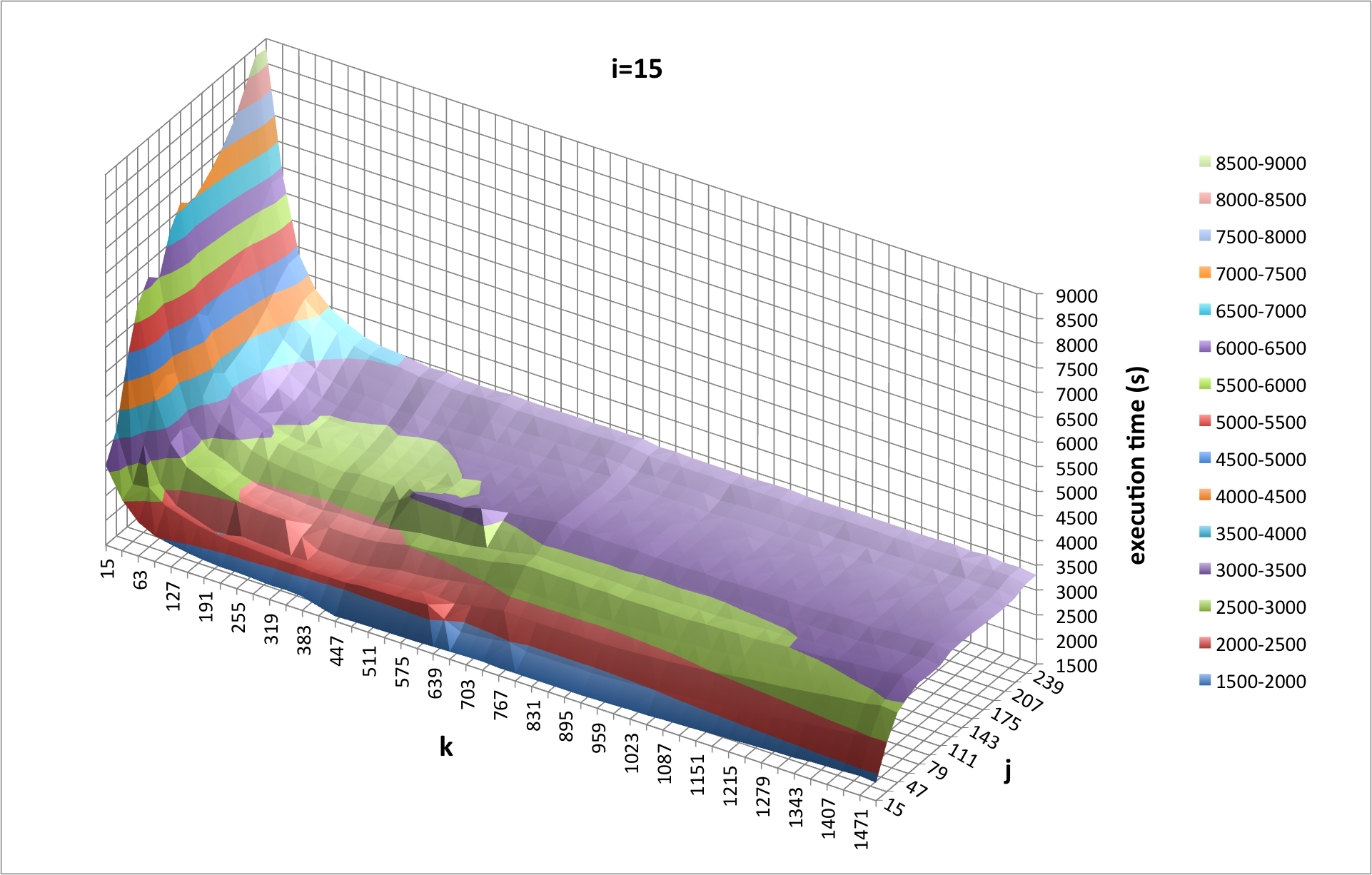}
\caption{\small Execution times of the stencil application as the blocksizes varies for i = 15}
\label{i153D}
\end{figure}

\begin{figure}[h]
\centering
\includegraphics[scale=0.37]{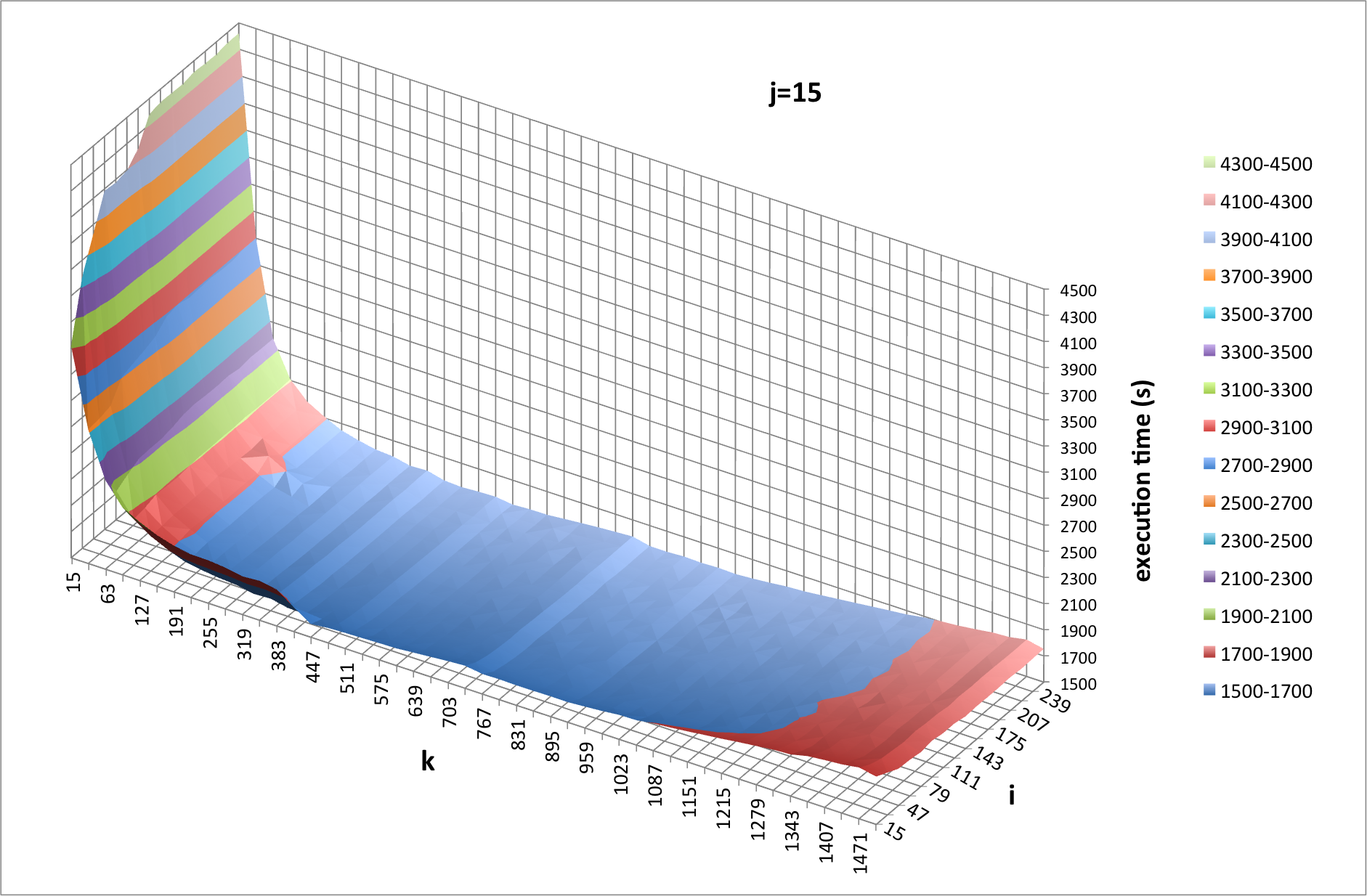}
\caption{\small Execution times of the stencil application as the blocksizes varies for j = 15}
\label{j153D}
\end{figure}

Figures~\ref{i153D} and~\ref{j153D}  presents the performance of the application according to the various blocksizes. Figure~\ref{i153D} shows that there is a large and conflicting variation on the application performance  as the values of $k$ and $j$ increase. While the performance of the application always diminishes as the value of $j$ increases, there is an exponential decay as the value of $k$ increases, but only for small values of $k$, more specifically until $k = 255$. After this point, a slight  deterioration on the execution time is produced until the best value is reached, when a worsening can be seen.

Note that   Figure~\ref{i153D} considers only $i=15$, since the behaviour of the application for an increasing value of $i$ is very similar. 
 This can be seen in Figure~\ref{j153D}, where $j$ is set to $15$ (one of the best values for this coordinate) while the values of $i$ and $k$ increases. One can see that the behaviour of the execution time as the value of $k$ increases is the same as the one showed in the previous figure. It is also important to remark that there exists a variety of very efficient blocksizes which are concentrated in the  middle part of the curve, roughly  between $k = 383$ and $1023$. 

In order to analyse the quality of the blocksize produced by $MWMB$ algorithm,  a comparison of the execution times between the best blocksize found through the exhaustive search and the one found by $MWMB$ algorithm is carried out. One of the best blocksizes found by the exhaustive search was for $i=183, j=15, k= 511$. In the case of the $MWMB$ algorithm for the same instance, the  blocksize found  was $47 \times 15 \times 511$. 

\begin{figure}[h]
\centering
\includegraphics[scale=0.7]{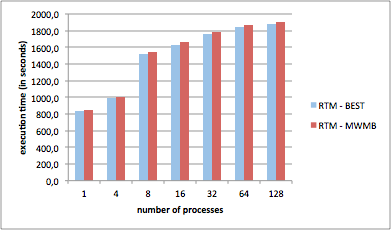}
\caption{\small Comparison of the execution time of the RTM program with the best the $MWMB$ blocksizes}
\label{graficobestxMWMB}
\end{figure}

As a matter of comparison, the  RTM program was executed for a process workload of $250 \times 250 \times 1500$ with both  blocksizes values ($i=183, j=15, k= 511$ and $i=47, j= 15, k= 511$) with an increasing  number of processes.  
 The results, shown in Figure~\ref{graficobestxMWMB}, validate the effectiveness of algorithm $MWMB$, as the loss of performance was insignificant. 
In the worst case, the execution with the $MWMB$ blocksize was only 2.4\% longer than the one with best blocksize, while for executions with  the largest number of processes, there was  a loss of only 1.3\%. Bear in mind that to identify the best block size via the exhaustive search,  thousands of experiments were  conducted during practically two months. Therefore, the quality of solution loss of $MWMB$ is nothing comparable with the time spent to obtain it by the exhaustive search strategy. 

\subsection{A metric to evaluate the quality of the blocksize}

The blocksizes produced by $MWMB$ showed to be efficient, providing execution times very close to the one obtained through an exhaustive search. However, this evaluation and assurance of the quality was possible due to a large number of executions to identify the best blocking value.  

%It would be useful to have means to measure the quality without the need for an exhaustively execute for the search of the best blocksize. 
%Thus, for a scientist to spend so many time to obtain a good solution is, in most cases, impracticable.   

%Thereby, it would be useful to a scientist to know (or at least have an idea) if the blocksize provided by an algorithm is efficient or not. It would help even more if the scientist could know how far from a good solution the blocksize is. Thus, this section presents a simple way of evaluating the efficiency of a blocksize produced by an algorithm.

Table~\ref{minActualTimes} presents results for the instances $200\times 200\times 800$, $ 300 \times 300\times 800$ and $250\times 250\times 1500$ considering $OB$, $MMMB$, $MWMB$ and $MAMB$. The table shows one of the best  blocksize produced by the corresponding methodology, the total execution time of the RTM application for the given blocksize, denoted as $Total$, the $MinTime$ and $ActualTime$ times, as already defined. 

% As a matter of fact, the time that will define the performance of the code is the {\em actual time} since it represents the main execution of  the  RTM code on the available  shared memory cores on each machine of the target cluster. 

\begin{table}[htb]
  \centering
  \caption{The impact of the comparison between Minimum and Actual times ($MinTime$ and $ActualTime$) on the execution time of the original parallel application ($Total$)}
  
  \label{minActualTimes}
  \begin{scriptsize}
\begin{tabular}{|c|c|c|c|c|c|}
  \hline
  % after \\: \hline or \cline{col1-col2} \cline{col3-col4} ...
  problem size   								& algorithm & block value & $Total$   &   $MinTime$  	&  $ActualTime$  \\  \hline
  {}       											& OB        & 159 209 175 & 945.6     &   3.09       	&  7.50  \\
  {}         										& MMMB 			& 159 209 191 & 944.2     &   2.59       	&  7.71  \\
  $200 \times 200 \times 800$  	& MWMB  		& 15 15 143   & 628.8     &   4.00       	&  3.99  \\
  {}         										& MAMB 			& 175 15 127  & 692.7     &   3.65       	&  4.60  
   \\ \hline
  {}           									& OB 				& 15 15 127   & 1296.2    &   8.74       	&  9.38 \\
  {}          									& MMMB 			& 175 15 127  & 1452.5    &   6.35        &  10.82 \\
  $300 \times 300 \times 800$ 	& MWMB 			& 15 15 159   & 1295.1    &   9.40        &  9.35 \\
  {}           									& MAMB 			& 175 15 127  & 1414.5    &   8.77        &  10.85 
  \\ \hline
  {}          									& OB        & 159 259 607 & 3646.3    &  14.67        &  33.27 \\
  {}          									& MMMB  		& 175 15 127  & 2146.2    &  12.65        &  16.20 \\
  $250 \times 250 \times 1500$ 	& MWMB 			& 47 15 495  	& 1949.0    &  13.60        &  13.57 \\
  {}          									& MAMB 			& 175 15 127  & 2121.3    &  12.91        &  15.90 \\
  \hline
\end{tabular}
\end{scriptsize} 
\end{table}

Based on a series of experiments carried out in this work, the comparison between  {\em minimal} and {\em actual times} is proposed as a metric of the quality of the produced results. One aspect concluded  is that when both {\em minimal} and {\em actual } times are close, the allocation of the shared memory during the {\em selection} phase reflects the allocation during the {\em verification} phase. Therefore,  
when the difference between  the {\em MinTime} and {\em ActualTime} is negligible,   the algorithm found a blocksize that leads to a faster execution when compared with  the execution times of all the others blocksizes. In other words, the closer are these two metrics, more efficient are the results.

As seen in Table~\ref{minActualTimes}, when {\em MinTime} and {\em ActualTime} are not near,   the $Total$ time is much higher than when both metric are close, since in this case, the new methodology succeeded in obtaining a blocksize in a specific  condition that does  reflect the actual memory hierarchy utilisation. For instance, the percentage difference of {\em MinTime} and {\em ActualTime}  for the $250 \times 250 \times 1500$ problem size of the algorithms $OB$, $MMMB$, $MWMB$ and $MAMB$ were 126.8\%, 28.1\%, -0.2\% and 23.2\%, respectively, being the actual execution times produced their respective blocksizes  $3,646.3$s, $2,146.2$s, $1,949.0$s and $2,121.3$s.

%Besides that, if there is an {\em actual time} produced by an efficient blocksize to be used as a comparison, the scientist can use this to evaluate the performance of his algorithm.  

Another aspect important to point out is that, the actual execution time (the verification phase execution time, $ActualTime$) of the RTM kernel having as input the blocksizes identified by $OB$, $MMMB$ and $MAMB$  were 145.2\%, 19.4\% and 17.2\% larger than the one associated with the blocksize identified by $MWMB$, respectively. Having in hands these blocksizes  and executing the original RTM parallel code leads to the execution  time  $Total$, which  were 87\%, 10.1\% and 8.8\% worse than the one executed with the blocksize produced by $MWMB$, respectively. Although, the  $ActualTime$  is not the same as $Total$, since they depict the execution of different number of iterations of the RTM code,  it can truly represents the performance of the four algorithms. 

%The algorithm $MWMB$ is the most efficient and the loss in performance are proportional in both metrics. 

%However, in most cases, scientists will not have a efficient blocksize to compare. In this case, a good metric to evaluate a blocksize is a comparison of the {\em MinTime} and {\em ActualTime} produced by the algorithm. 

%As can be seen in Table~\ref{minActualTimes}, if the difference among these times are very close to each other it means that the algorithm have found an efficient blocksize. On the other hand, the efficiency of the blocksize diminishes in the same proportion as the difference among these times increase. When the {\em MinTime} and {\em ActualTime} are really close it means that the algorithm found a blocksize that were faster than all the other blocksizes tested and also that this blocksize maximizes the memory hierarchy efficiency. Otherwise, if {\em MinTime} is smaller than {\em ActualTime} it represents that the algorithm obtained a blocksize in a specific memory condition that does not uses the memory hierarchy efficiently. For instance, the percentage difference of {\em MinTime} and {\em ActualTime}  for the $250 \times 250 \times 1500$ problem size of the algorithms $OB$, $MMMB$, $MWMB$ and $MAMB$ were 126.8\%, 28.1\%, -0.2\% and 23.2\%, respectively. While the execution times produced by the blocksizes obtained by the algorithms $OB$, $MMMB$, $MWMB$ and $MAMB$ were 3646.3s, 2146.2s, 1949.0s and 2121.3s, respectively.

% QTD de combina›es testadas
\subsection{On the evaluation of the number of combinations analysed during the selection phase}

%The analysis of the quality of the blocksize carried out in subsection~\ref{quality}  showed that the algorithm $MWMB$ produces efficient blocksizes. Moreover, it also highlighted that there is not only one but many efficient blocksizes. In other words, there were several different combinations of values for each domain that provides high quality blocksizes. 

%Another important aspect that should be pointed out is that the new technique proposed in this work in which all the cores of the same node execute the same combinations of blocksizes in the same order will increase approximately $C$ times, where $C$ is the number of cores, the execution time of the algorithm when compared with the original algorithm ($OB$), if both execute the same amount of combinations (until now all executions of each algorithm tested 8,000 combinations). 

This section investigates the  quality of the blocksize obtained by $MWMB$ algorithm as the number $nC$ of combinations tested during the {\em selection phase} increases. For doing so, the algorithm $MWMB$ is adapted to evaluate all the problem size domain according to the number of combinations to be tested. Being a 3D problem, the space domain is divided in the three dimensions and therefore,  the combinations to be tested are chosen based on the number of divisions of the problem size. Let the size of each $l$ dimension of the 3D problem be $S_l$. This $S_l$ will be divided in  $P$ parts . Therefore, the values to be tested in the $l$-dimension are 
\[ b_l = \langle 1,   \frac{S_l}{P-1} + 1, \frac{2S_l}{P-1} + 1, \frac{3S_l}{P-1} + 1, \ldots , \frac{(P-1)S_l}{P-1} + 1\rangle\]
 Therefore, the total number of blocksize values  tested are $nC = \mid b_i \mid\times\mid b_j \mid\times\mid b_k\mid $.

Table~\ref{combinacoes} shows the results of this experiment for three  values of $P$ (parts)  of the problem domain: $P = 5, 10 , 20$, resulting in three set of combinations ($nC = 125, 1,000$ and $8,000$). As an example, for the problem size $200 \times 200 \times 800$, the values tested for the $i$-dimension when $P=5$ are the combination of : $b_i = \langle 1, 51, 101, 151,  200 \rangle $, $b_j = \langle 1, 51, 101, 151,  200 \rangle $ and $b_k = \langle 1, 201, 401, 601,  800 \rangle $, a total of $nC = 125$ combinations (note that the number of parts being evaluated are the same for all of the three dimensions,  but this is not mandatory). 

For each one of the three sets of combinations of each problem size instance, it was carried out a series of 10 executions in 8 nodes from Oscar of  the  RTM code with the produced best blocksize value, and a comparison with the execution time of the selection phase under  the $MWMB$ procedure considering all the $nC$ combinations of block values. The column {\bf Total} represents the execution time of  the  RTM parallel code  using the blocksize $b_{MWMB}$  calculated by algorithm $MWMB$, while  column $Total_{MWMB}$ shows the necessary time for $MWMB$ to find $b_{MWMB}$, i.e.\ all the iterations necessary for  {\em selection phase} considering all the $nC$ combinations added to the {\em verification phase}.

As can be seen in Table~\ref{combinacoes}, the  time $Total_{MWMB}$ increases proportionally to the  number of combinations, as expected. In the case of the $300\times 300\times 800$ problem size instance, while the total number of combinations tested increased from $125$  to $1,000$ and $8,000$ (8 and 64 times more, respectively), the execution time of the selection phase $Total_{MWMB}$ increased 619\% and 680\%, respectively. On the other hand, the gain in performance as the number of combinations tested increased was quite modest. For instance, considering the average results for the $300\times 300\times 800$ problem size, while $MWMB$ was executed practically 60 times longer evaluating 8,000 blocksize combinations than the time spent to evaluate 125 combinations, the $Total$ time spent to execute the original RTM parallel code with the best blocksize out of 8,000 combinations improved no more than 7\%. 

\begin{table}[htb]
  \centering
\caption{Total execution time of the RTM code (10 executions) using the blocksize $b_{MWMB}$ produced by $MWMB$ and $Total_{MWMB}$ spent by the procedure to identify $b_{MWMB}$  for $nC$ combinations.}
  \begin{scriptsize}
\begin{tabular}{|c|c|c|c|c|c|c|}
  \hline
{} & \multicolumn{6}{c|}{Total of combinations tested} 	    \\ \cline{1-7}

  {} & \multicolumn{2}{c|}{ $P=5$ ($nB = 125$)} 	& \multicolumn{2}{c|}{$P=10$ ($nB = 1,000$) } & \multicolumn{2}{c|}{$P=20$ ($nB = 8,000$)}        \\ \cline{1-7}

  Size     & $Total $   & $Total_{MWMB}$    & $Total $    & $Total_{MWMB}$  & $Total $    & $Total_{MWMB}$    \\  \hline
  {}           & 658.2  & 193.3       &   583.7          &  1422.2  &  568.8 & 10972.8\\
  {}           & 660.8  & 192.6       &   593.4          &  1426.6  &  575.3 &   10970,0\\
  {$200$}           & 657.5  & 192.4       &   582.5          &  1427.8  &   655.1 &10956.7\\
  {$\times$}           & 656.2  & 191.2       &   577.3          &  1427.4  &   606.7  & 10973.4\\
  $200$  & 662.0  & 194.2       &   585.7          &  1443.5 & 578.2 & 10986.4 \\
  {$\times$}           & 656.0  & 192.3       &   592.2          &  1429.3  & 580.0  & 11122,1\\
  {$800$}           & 654.1  & 193.3       &   582.7          &  1428.4  &  585.0 &11074,4\\
  {}           & 655.8  & 191.5      &   578.4          &  1432.5  & 574.9   &  11094,2\\
  {}           & 653.7  & 195.3       &   588.4          &  1431.2  &   577.3 &  11064,9\\
  {}           & 654.3  &195.7       &   578.5          &  1435.8 &  599.0    & 11055,4
  \\ \hline
  {A}           & {656.9}  & {193.2}       &  {584.3}          &  {1430.5}  & {590.0}    & {11027.0}
  
  \\ \hline
  
  {}           & 1370.1    & 541.7      &  1298.4     &  3908.4 & 1196.5 & 30407.2 \\
  {}           & 1359.8    & 542.1      &   1308.9    &  3899.1 & 1235.5 & 30869.8 \\
  {$300$}           & 1366.6    & 543.1      &   1282.4    &  3899.3 & 1192.1 & 30316.1\\
  {$\times$}           & 1369.4    & 541.6      &   1300.3    &  3901.1 & 1187.7 & 30329.6 \\
  $300$  & 1379.2   & 540.1      &   1279.1       &  3901.1 &1203.6  & 30407.5\\
  {$\times$}           & 1381.1   & 543.3      &   1308.9        &  3897.5 & 1238.1 & 30760.6\\
  {$800$}           & 1394.7   & 544.1      &   1267.1        &  3898.8 & 1185.5 & 30369.8 \\
  {}           & 1388.2   &543.2      &   1311.6        &  3898.4 & 1211.8 &30319.6\\
  {}           & 1364.3  & 540.9      &   1266.3       &  3897.3 & 1209.8&30265.6\\
  {}           & 1377.6    & 544.0      &   1317.2        &  3888.8 & 1169.8 &30236.0
  \\ \hline
  
  {A}           & {1375.1}  & {542.4}       &  {1294.0}          &  {3899.0}  & {1203.0}    & {30428.2}
  
  \\ \hline
  
  {}           & 2033.2  & 1074.2         &    1906.5      & 7826.6  & 1898.8  & 61986.5\\
  {}           & 2076.8  & 1082.4         &    1902.2      & 7863.9  &  1856.1 & 61327.9\\
  {$250$}           & 2075.6  & 1080.4         &    1926.9      & 7825.0  &  1885.7 & 61867,1\\
  {$\times$}           & 2057.6  & 1059.9         &    1900.8      & 7850.9 &   1944.7& 61348.8\\
  $250$    & 2060.4  & 1077.9      &  1934.8       & 7851.7  & 1944.3 & 61319.7\\
  {$\times$}           & 2056.6 	& 1079.8      &     1920.7       & 7813.4  & 1965.3 & 61655.8\\
  {$1500$}           & 2066.7     & 1085.0      &      1935.6      & 7844.1   & 1947.9 & 61523.3\\
  {}           & 2024.7     & 1089.7      &      1912.4      & 7864.9  & 1847.7 & 61523.5\\
  {}           & 2039.5 &  1074.4         &      1892.8      & 7815.7  & 1861.3 & 61732.1\\
  {}           & 2031.0 & 1079.4          &       1922.6     & 7842.4 &  1859.7 & 61604.8\\
  \hline
  
  {A}           & {2052.2}  & {1078.3}       &  {1915.5}          &  {7839.9}  & {1901.1}    & {61588.9}
  
  \\ \hline
  
\end{tabular}
\label{combinacoes}
\end{scriptsize}
\end{table}

The performance for the other two problem sizes were similar, although for the $200\times 200\times 800$ problem size instance,  the average $Total$ time associated with the blocksize found when $nC = 8,000$  were slightly worse than the one associated with the blockizesize for $nC = 1,000$. However, the smallest $Total$ time amongst all the 10 executions for this instance for each number of combinations was 568.8 seconds, which was associated with $nC = 8,000$. 

\begin{figure}[htb]
\centering
\includegraphics[scale=0.47]{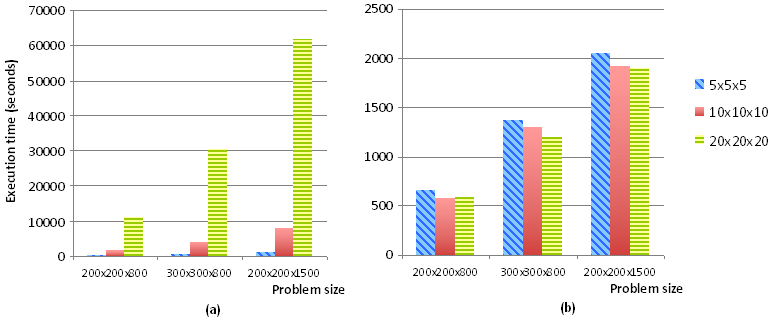}
\caption{\small (a) Average execution time to find the blocksize by $MWMB$. (b) Total execution time produced by RTM using the blocksize calculated}
\label{BarraExecTime}
\end{figure}

Figure~\ref{BarraExecTime}(b) clearly shows the slight performance gain while $Total_{MWMB}$  (the execution time of the {\em selection phase}) drastically increases as  the number of combinations $nC$  grows (Figure~\ref{BarraExecTime}(a)).

When executing RTM application for a given instance for the first time, it is necessary to identify an efficient blocksize value, as already known. Suppose that the application will be executed only once. In this case, the overall execution time would be  $Total_{MWMB}  (nC)+  Total $, i.e.\ is the execution time taken by  $MWMB$ to identify the best blocksize $b_{nC}$ amongst $nC$ combinations plus the execution time of RTM application using $b_{nC}$. Considering the  best executions times (the smallest $Total$ for the 10 executions) for three distinct set of combinations ($nC = 125, 1,000$ and $8,000$) performed for the  instance $250 \times 250 \times 1500$, 2,024.7s, 1,892.8s and 1,847.7s and its respective $Total_{MWMB}(nC)$ 1,089.7s, 7,815.7s and 61,523.5s,  it would be necessary for the RTM application using $b_{125}$ to be executed 51 times more than RTM application using $b_{1000}$.

% would provide a total time (execution time + algorithm time) more efficient. Conversely,  the application would take approximately more than 691,301 seconds (192 hours) for  $125$ combinations so that the blocksize calculated for $8000$ combinations would provide a better performance.

%\textcolor{red}{The above paragraph is very confusing. Observing the above table we can calculate Total+TotalMinTime: \\
%For $nC = 125$: 
%$2024,7 +  1089,7= 3114,4 seg = 51 min < 1 h $ \\
%For $nC = 1000$:
%$1892,8 + 7815,7 = 9708,5 s = 161min =  < 3h $\\
%For $nC = 8000$:
%$1847,7 + 61523,4 = 63371,25 s = 1056,18min =  < 17h $\\ \\
%Thus, whu these 29 hours for 125 combinations. 192 hours for .... Maybe, I do not understand the text.....
%}

%In practical terms, the actual scientist code will run a larger number of iterations than what was executed in the experiments carried out here, taking hours or days to be executed.

Thus, when choosing the number of combinations to be tested, the scientist should take into account the average execution time of the application and the number of times it will be executed. For applications that will run for a few hours, a small number of combinations is enough to provide a good blocksize and the algorithm will not be so intrusive. However, if the application is due to be running several times for many hours or days, a larger number of combinations should be tested.  

\section{Related Work}

An investigation of the performance of evolving memory systems features, such as large on-chip caches, automatic prefetch, and the growing distance to main memory on 3D stencil computations was presented in~\cite{Kamil:2005}. The main observation is that improving cache reuse is no longer the dominant factor to consider in optimising these computations. In particular, the authors considered that streaming memory accesses are increasingly important because they engage software and hardware prefetch mechanisms that are essential to memory performance on modern microprocessors.

In~\cite{ccgrid2007} the impact of multicore architecture was analysed based on a case study with a Intel Dual-Core system. The main conclusion was that in these machines the intra-node communication has a significant impact and cache and memory contention can be a bottleneck. Thus, they state that the applications should be multicore aware to overcome these problems. The conclusions were made based on benchmarks. However,  the study was poor in the sense that at most four nodes with two cores each were used.

In~\cite{scStencil2008}, a suite of  optimisation mechanisms were evaluated considering a set of modern architectures.  They developed an auto-tuning architecture aware environment similar to ATLAS and OSKI.  The approach provided good performance across a variety architectural configuration. The second component of an auto-tuner is the auto-tuning benchmark that searches the parameter space through a combination of explicit search for global maxima with heuristics for constraining the search space. At completion, the auto-tuner reports both peak performance and the optimal parameters.

In~\cite{Nguyen:2010}, a 3.5D blocking algorithm based on a 2.5D spatial blocking together with a temporal blocking on the input grid into a on-chip memory is proposed to increase the execution performance for Stencil Computations. The 2.5D spatial blocking blocks into two dimensions and then, outflows through the third dimension, in such a way that~\cite{Nguyen:2010} advocates that the improvement on the on-chip memory utilization is due to the reduction on memory bandwidth usage. The temporal blocking executes multiple time-steps of the stencil computation, and thus data re-used.  The performance of the algorithm is compared with other algorithms and it is faster for both CPU and GPU implementations, being almost linearly scalable. However, different from the work proposed here, the work does not infer how the dimensions of the blocking are calculated. Also, a good performance is only achieved if there is enough cache capacity to hold the blocked data. 

%An analysis of impact of executing a real RTM code in a multicore system is presented in~\cite{eScience2011}, evaluating four versions of the code. Based on this analyses, the code is optimised for multicore systems utilising two techniques: reduction of the number of instructions through loop transformations; a better use of memory hierarchy by choosing an adequate blocksize.    

In~\cite{Byun2012} the auto-tuning framework pOSKI for sparse linear algebra kernels on multicore systems is proposed as an extension of a previous work devised for cache-based superscalar uniprocessors. pOSKI applies a block compressed sparse row since it can achieve reasonable performance when proper register block size is selected. Both off-line and runtime auto-tuning mechanisms are conducted by pOSKI.  The off-line auto-tuning considers a set of tunable parameters collects benchmarking data sparse matrix for a given architecture and compiler. Finally it selects the best implementation for each register block size. The authors advocate that although expensive, the off-line auto-tuning is effective since it is done once per architecture and compiler before running the application.  The runtime auto-tuning is performed only after the matrix nonzero locations are known. Firstly, the sparse matrix is partitioned in sub-matrices compounded of consecutive row blocks with similar number of non-zeros per block, each one executed by a thread. However, due to the use of different register block sizes, load unbalance can occur. Actually, a heuristic performs the register block size choice, which is only based on the number of non-zero elements. Also, the choice of an optimal number of cores is part of the auto-nuning problem. Currently, pOSKI just uses all available cores that are provided, since to identify the right number of cores is costly. Still, work has to be done on how to find a number of cores (not necessarily all the cores of a machine) in order to achieve good performance.

Versions of the Stencil kernel have been proposed and implemented not only on conventional CPU multicore processors, but also on FPGAs~\cite{Nemeth2008} and GPGPU ~\cite{Micikevicius:2009,ICCS2011,Tao:2012}. General CPU systems are the most flexible, allowing a variety of optimizations and algorithmic manipulations. Furthermore, the majority of the RTM programs available are implemented in FORTRAN or C, using MPI or OpenMP, and are portable to these high performance platforms.  Also, the CPU implementations allows to execute large domain problems with higher orders(and therefore, more points can be considered).

The work presented here is different from others, including the ones related to tiling techniques as presented in~\cite{Abella02,Rivera2000},  in the sense that it proposes  an algorithm to optimise the use of memory hierarchy by 3D domain problems.

%------------------------------------------------------------------------
\vspace{-0.1cm}
\section{Concluding Remarks}

Although the new CPU multicore machines are able to provide more computational power to the scientists, the performance of the 3D stencil applications depends on an efficient utilisation of the memory hierarchy of these machines. The tiling or blocking technique can be used to obtain an improvement of the data locality and thus a more efficient execution.

This work proposes a simple algorithm to produce an effective blocksize to be used in 3D stencil computations such as the RTM code or any problem where processes compete for shared cache space. More important, a new approach to utilise the multicore machines are introduced in this algorithm and the results confirms the advantages of this technique. The benefits of this new algorithm were highlighted through a series of experiments and when compared with the best block obtained through exhaustive search for an specific problem size it was only 2\% slower.  Besides, experiments showed that a small number of combinations can produce very effective blocksizes and a simple way of evaluating the efficiency of the blocksize was presented.

%-------------------------------------------------------------------------
\bibliographystyle{plain}
\bibliography{refRTMJournal}

\end{document}